\documentclass{article}

\usepackage[margin=1in]{geometry}
\usepackage{times}
\usepackage{helvet} 
\usepackage{courier}   \usepackage{amsmath}
\usepackage{amssymb}
\usepackage[utf8]{inputenc} 
\usepackage[T1]{fontenc}    
\usepackage{hyperref}       
\usepackage{url}            
\usepackage{booktabs}       
\usepackage{amsfonts}       
\usepackage{nicefrac}       
\usepackage{microtype}      
\usepackage{natbib}

\usepackage{graphicx}

\usepackage{caption}
\usepackage{subcaption}
\usepackage[monochrome]{xcolor}
\usepackage{algorithm}
\usepackage{algpseudocode}
\usepackage{bbm}
\usepackage{enumitem}

\title{A Contextual Bandits Approach for Personalization of Hand Gesture Recognition}

\author{
 Duke Lin \\
  Reality Labs, Meta \\
 dukelin95@gmail.com \\
   \and
Michael Paskett \\
  Reality Labs, Meta \\
 michaelpaskett@meta.com \\
  \and
Ying Yang \\
  Reality Labs, Meta \\
  ying.yang@meta.com \\
}

\begin{document}

\maketitle

\begin{abstract}

In human-computer interaction applications like hand gesture recognition, supervised learning models are often trained on a large population of users to achieve high task accuracy. However, due to individual variability in sensor signals and user behavior, static models may not provide optimal performance for all users. Personalizing pretrained models via calibration--collecting labeled data from each user--can improve performance but introduces user friction and struggles with limited data. To overcome these issues, we propose a calibrationless longitudinal personalization method: a contextual multi-arm bandit (MAB) algorithm combined with a pretrained neural network for gesture recognition. This reinforcement-learning-style approach enables personalization using binary reward signals, either user-provided or inferred by the system.

We validated this method in a user study. Participants wore a surface electromyography (sEMG) device and played multiple rounds of a 2-D navigation game using six hand gestures. In the session, they completed a baseline round and then a round with our algorithm; in the second session, they played another round with our algorithm. Our approach led to a significant reduction in users' average false negative rate by 0.113 from the initial to the final round, with further decreases between sessions. Average precision also trended upward (by 0.139) from the start to end of a round, continuing in the next session. Notably, some users who could not complete the game with the baseline model succeeded with our contextual MAB model. In summary, our

\end{abstract}

\begin{figure}[t]
  \includegraphics[width=\textwidth]{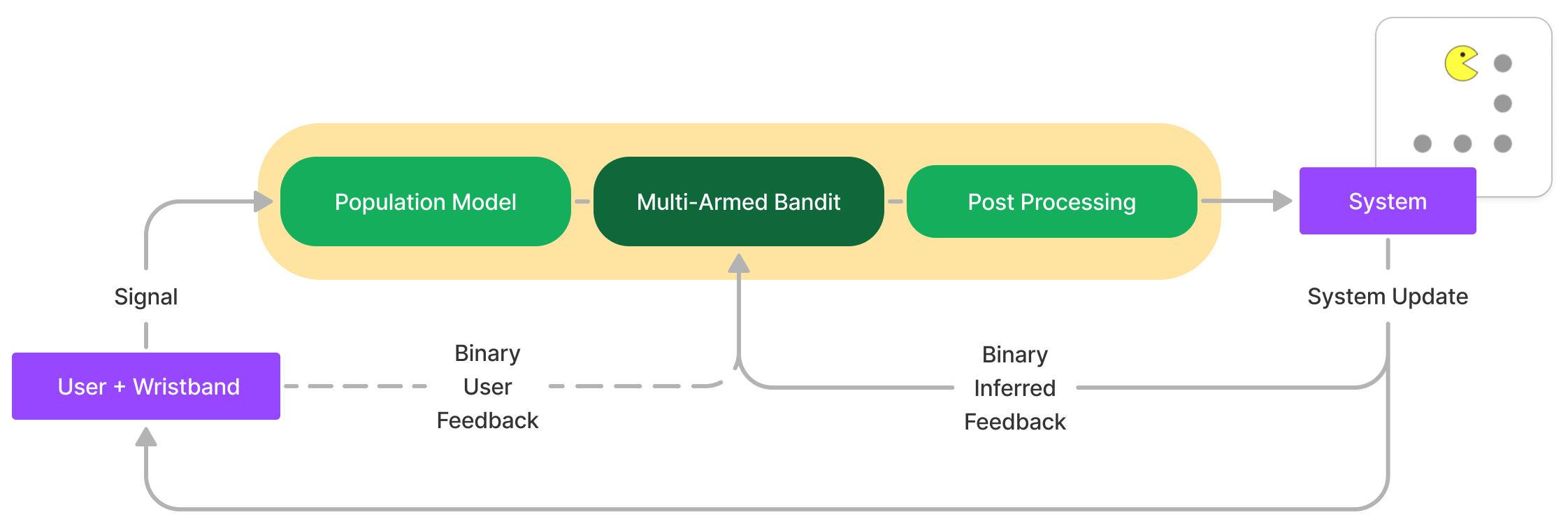}
  \caption{A participant's signals are decoded by the proposed algorithm to play a navigation game. Using reward signals from the user and the system, the model enables real-time personalization for the user over time.}
  \label{fig:teaser}
\end{figure}

\section{Introduction}

Hand gestures, such as directional thumb swipes for navigation with index and thumb taps for selection, provide an intuitive interface for new computer interactions. 
Decoding hand gestures from wrist-worn devices providing signals from inertial measurement units (IMU) and surface electromyography (sEMG) electrodes is an appealing method for developing new interactions (\cite{xu2022enabling} and \cite{ctrl2024generic}). 
Recent developments in sEMG decoding have shown that with a large dataset and supervised learning methods, gesture recognition models can generalize to new users (\cite{ctrl2024generic}). 
However, the model is ultimately static and does not work perfectly for all users at all times. 
Thus, to improve real-time performance and combat individual variability of signals, there is a need to supplement the performance of pre-trained models with personalization – improving model performance with user-specific data.

One method of acquiring user specific training data is a calibration process either during the set up of a new device or at a user-driven cadence.
By having the user provide explicit samples of each gesture, the underlying model can then be fine-tuned on the additional data as a means of personalization. 
However, this step incurs additional overhead to the user and can introduce additional errors if the user does not follow the calibration process perfectly. 
Moreover, the model may also face challenges in over-fitting from fine-tuning on a small set of calibration data.

In an effort to improve overall user experience and reduce users' cognitive load, we seek to extract implicit learning signals during regular device usage.
These learning signals can be sparse and binary in nature. 
In practice, this approach can live in the background providing personalization at predefined cadences. 
For example, since we rely on sparse reward signals from the user's time with the device, a cadence can be based on a threshold with respect to the amount of rewards accumulated.
By leveraging the recent advances in reinforcement learning for real-time model adaptation, we seek to enable longitudinal online personalization with a contextual multi-arm bandit (MAB) algorithm as the final layer of a population-trained gesture recognition model. 
This algorithm updates its parameters based on reward signals from the system as well as optional rewards from the user.
With a learned linear mapping from the population model embeddings to a reward estimate for each gesture, the model is simple to update and yields improvements in real-time.
To our knowledge, this is a novel approach to online longitudinal personalization for sEMG-based gesture recognition.

We applied our algorithm to sEMG model, enabled by the Meta sEMG Research Device (sEMG-RD), a dry-electrode, multi-channel wristband (\cite{ctrl2024generic}). 
The wireless device features a simple form factor, enabling easy donning and doffing, and facilitating broader usage compared to most existing EMG systems.
The wristband possesses 16 channels of bipolar sEMG at 2 kHz, evenly spaced around the wristband.
The signals are processed and decoded into gestures for computer interaction via Bluetooth connection.
A pre-trained gesture detection neural network maps sEMG and IMU inputs to an intermediate embedding for our algorithm.

Utilizing this device and a 2-D navigation game, we conducted a multi-session user study to demonstrate that our method boosts model performance over time. 
In the game, six gestures —  directional thumb swipes, index tap and thumb tap  — were used to control the character and navigate a pre-defined trajectory. 
We can infer reward from either the progression of the character or the feedback provided by the participant.
With real-time updates of the contextual bandits, we measured significant increases in precision scores by comparing the score from the end of a round to the beginning of a round as well as a decrease in false negative rate over time.
With multiple sessions, we were able to establish continued improvement in between session, demonstrating the models capability of longitudinal improvement. 

In summary, our major contributions are:
\begin{itemize} 
    \item a novel approach to improve pre-trained neural network recognition models using binary reward signals without requiring an explicit user calibration.
    \item a real-time framework of updating the model and personalizing over time.
    \item a user study leveraging a 2D navigation game experience to demonstrate model improvement.
\end{itemize}

\section{Related Work}

In this section, we briefly summarize related work on improving pre-trained models for individual users, a process often referred to as personalization. 
These methods have been applied to various settings such as human-machine interfaces, gesture and pose recognition, and recommendation systems.
Similar techniques are also used to generalize a pre-trained model to new tasks --- for example, adding new customized gestures to an existing gesture recognition model (\cite{xu2022enabling}).

\subsection{Supervised and Unsupervised Approaches} 
Retraining or fine-tuning a model with supervision is an intuitive method for enhancing model performance.
In this paradigm, data and labels are required from the user, generally collected as a calibration dataset.
This approach is widely applied in brain-computer interfaces, where neural or electromyographic signals serve as input, and a relatively simple model (e.g., linear mapping) is used to map the input signal to the position of a 1D/2D cursor, enabling users to move it to target locations.

Studies such as \cite{madduri2024predicting} and \cite{despradel2024enabling} have demonstrated closed-loop solutions to the model parameters can be obtained within a short amount of time by leveraging the target position as labels.
In these cases, real-time calibration of the personalized model is performed while users control a cursor. 

For higher-capacity models, often only a subset of the model's parameters are fine-tuned from the calibration data from new users or new tasks. 
In such cases, researchers seek to reduce the amount of calibration data required by increasing data efficiency, and introducing minimal user friction through well-designed interfaces.
For example, \cite{xu2022enabling} focused on recognizing gestures from inertial measurement units (IMUs) in the Apple Watch, leveraging a carefully designed user interface to obtain IMU data and gesture labels for user-customized gestures.
Data efficiency was improved by employing a delta-encoder to supplement the size and diversity of the training data.

To enhance data-efficiency during personalization, other works in this area leverage meta-learning techniques. 
For example, \cite{joshi2017personalizing} used a hierarchical Bayesian representation of the user-specific model parameters with a common prior, which is learned across training users and infers the best parameter from new user data and the prior.
\cite{akbari2021meta} and \cite{rahimian2021few} designed specific architectures in their neural network models, featuring shared parameters between users and a dedicated sub-network to customize user-specific model parameters given labeled data from a new user.
The dedicated sub-network can be trained using a set of existing users to capture the relationship from input data and the user-specific parameters that provide the best performance.
Subsequently, this learned mapping can generalize at personalization time for a new user.  

Besides existing work that improves calibration data efficiency, there are also lines of research in human-computer interaction applications that relax the requirement of calibration labels and aim to personalize with unlabeled input data.
For example, in gesture recognition, a semi-supervised learning approach was employed in \cite{shen2024boosting} where authors assumed a pre-trained high-capacity accurate model could label a large amount of input data, and use them to train the deployed small model (for real-time prediction) using a Connectionist Temporal Classification (CTC) loss, tolerating temporal uncertainty of the gesture events. Additionally , \cite{chen2023robust} exploited domain-adaptation techniques to map input data from new users to the distribution of the training users, demonstrating unsupervised personalization for hand gesture recognition. 

However, the semi-supervised and unsupervised learning approaches may still be limited compared to supervised learning.
In \cite{akbari2021meta}, the authors also explored only using input data without labels to train the sub-network for estimating decoding parameters for new users, and observed that the supervised approach (using both input data and labels) outperformed the unsupervised approach (using only input data without labels). 

\subsection{Reinforcement Learning Approaches}
Calibration data for supervised personalization can be hard to obtain and unsupervised techniques have their own limitations. Reinforcement learning approaches can be leveraged by determining meaningful reward signals from the system.

\cite{berman2024efficient} provides an example of using reinforcement learning to fine-tune the parameters of a musculoskeletal model, by generating a reward signal from hand pose error. 
In the Human-Computer Interaction (HCI) space, researchers have explored how users can interact with reinforcement learning algorithms without sacrificing the experience while showing improvements in user tasks. For instance, studies such as \cite{10.1145/3613905.3651059} and \cite{10.1145/3613904.3642063} demonstrate that the completion of a task can serve as a reward signal, enabling users to implicitly guide the learning process.

\cite{den2020reinforcement}, surveyed a wide range of reinforcement learning approaches for personalization, mostly for recommendation systems, but the ideas can be applied to wider applications.
One tractable approach is the multi-armed bandit (MAB) algorithm.
Multi-armed bandit algorithms are named after a hypothetical scenario in which a gambler playing multiple slot machines must determine which arm to pull to maximize their winnings.
Each machine provides a random reward from an unknown probability distribution, forcing the gambler to learn how to maximize their winnings while acquiring more information about each machine, a dilemma known as the exploration-exploitation trade-off.
MAB algorithms model a decision-making agent with internal estimates of the uncertainty and reward associated with each arm, enabling the agent to make informed decisions.
Contextual bandit algorithms extend the MAB formulation by incorporating contextual information or features related to each arm, improving the accuracy of uncertainty and reward estimates and enhancing downstream decision-making.

Linear Upper Confidence Bound (LinUCB) is a generalized contextual bandit algorithm that assumes the expected reward of an arm is linear to the features \cite{Li_2010}.
In addition, it leverages the confidence interval defined by the standard deviation of the expected reward of an arm as the uncertainty of the respective arm.
By modeling the uncertainty and reward as described, the arm with the highest upper confidence bound is pulled. 
Applied to gesture recognition, we can assume that predicting one gesture is equivalent to pulling one arm and assume that the population model embedding provides the necessary context.
After predicting a gesture (pulling an arm), we can observe the reward — whether the system advances or the user provides feedback.

\algnewcommand{\Initialize}[1]{%
  \State \textbf{Initialize:}
  \Statex \hspace*{\algorithmicindent}\parbox[t]{.8\linewidth}{\raggedright #1}
}

\algnewcommand{\Inputs}[1]{%
  \State \textbf{Inputs:}
  \Statex \hspace*{\algorithmicindent}\parbox[t]{.8\linewidth}{\raggedright #1}
}

\algnewcommand{\Outputs}[1]{%
  \State \textbf{Outputs:}
  \Statex \hspace*{\algorithmicindent}\parbox[t]{.8\linewidth}{\raggedright #1}
}

\algnewcommand{\IfThen}[2]{
  \State \algorithmicif\ #1\ \algorithmicthen\ #2
}
  
\begin{algorithm}[t]
  \caption{Main Algorithm}\label{alg:main}
  \begin{algorithmic}[1]
    
    \Inputs{$d$, $W$, $N$, $\alpha$}
    \Initialize{\strut $PopulationModel$, $System$ \\ $\boldsymbol{A}_{i} \gets \boldsymbol{I}_{d}$, $i=1,\ldots,N$ \\ $\boldsymbol{b}_{i} \gets \boldsymbol{0}_{dx1}$, $i=1,\ldots,N$}
    
    \For{$t = 0$ to $ T $}
        \State $\boldsymbol{e_{t}}$, $\boldsymbol{prob_{t}} \gets PopulationModel(\boldsymbol{signal_{t}})$
        \For{$i=1,\ldots,N$}
            \State $p_{t, i} \gets \boldsymbol{prob_{t}} + \boldsymbol{(A_i^{-1}b_i)^\intercal e_t} + \alpha \boldsymbol{\sqrt{e_t^\intercal A_i^{-1} e_t}}$
        \EndFor
        
        \State $arm_{t} \gets \arg\max_{i=1,\ldots,N} p_{t, i}$
        \State $class_{t} \gets PostProcess(arm_{t}$, $ Sum(\boldsymbol{prob_{t}}))$
        \State $r_{t} \gets System(class_{t})$

        \If {$ r_{t} \neq null $}
            \For{$j=1,\ldots,W$}
                \State $\boldsymbol{A}_{arm_{t-j}} \gets \boldsymbol{A}_{arm_{t-j}} + \boldsymbol{e}_{t-j}{\boldsymbol{e}_{t-j}}^{\boldsymbol{\intercal}} $  
                \State $\boldsymbol{b}_{arm_{t-j}} \gets \boldsymbol{b}_{arm_{t-j}} + r_{t}\boldsymbol{e}_{t-j}$ 
            \EndFor
        \EndIf
    \EndFor
  \end{algorithmic}
\end{algorithm}

\setlength{\textfloatsep}{1pt}
\begin{algorithm}[t]
  \caption{PostProcess}\label{alg:sub}
  \begin{algorithmic}[1]
    \Inputs{$arm$, $prob$}
    \Outputs{$class$}
    \Initialize{\strut$\tau_{b}$, $\tau_{e}$, $W$ \\$b^{a} \gets []$ \\ $b^{p} \gets []$}
    \State $b^{p} \gets prob$
    \If {$\max_{i=1,\ldots,W} b^{p}_{i} > \tau_{b}$}
        \State $b^{a} \gets  \boldsymbol{\mathbbm{1}_{x=arm}}$
        \State $\boldsymbol{m} \gets \frac{1}{W} \sum_{i=1}^{W}\boldsymbol{{b^{a}_{i}}}$
        \If {$\exists i \in \{1, \ldots, N\}: m_i > \tau_{e}$}
            \State $class \gets \arg\max_{i=1,\ldots,N} m_{i} - \tau_{e}$
            \State \textbf{return } $class$
        \EndIf
    \Else
        \State $b^{a} \gets []$
    \EndIf
  
  \end{algorithmic}
\end{algorithm}

\section{Methods}

\begin{figure*}
  \centering
  \includegraphics[width=\textwidth]{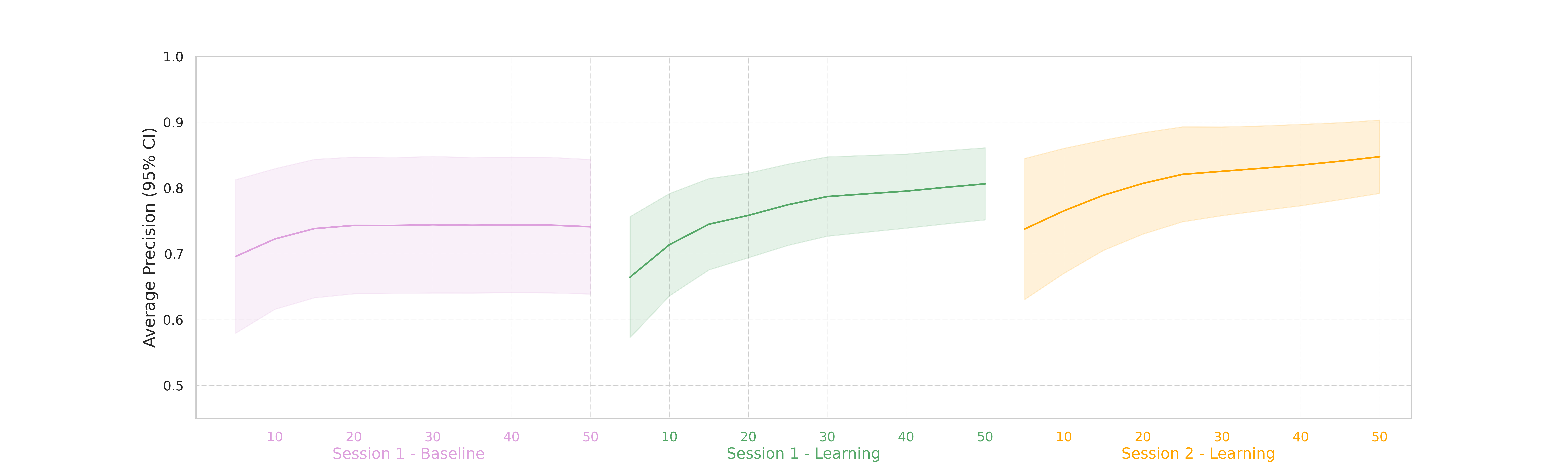}
  \caption{Average precision across all participants and gestures, where each data point is calculated after five trials. Compared to the baseline with no learning, the next two rounds with our algorithm show a positive trend of improvement for participants on average.}
  \captionsetup{justification=centering}
  \label{fig:average_performance}
\end{figure*}

\subsection{Contextual Multi-armed Bandit}
In our solution (Algorithm \ref{alg:main}), the sEMG and IMU signal from the device at time \(t\), denoted as \(\mathbf{signal_{t}}\), is first embedded by the pre-trained population model to obtain an embedding \(\mathbf{e_t}\) of \(d\)-dimension.
For each of the \(N\) gestures, we initialize an arm with a \(d\)-dimensional identity matrix, \(\mathbf{A_i}\), and a \(d\)-dimensional zero vector, \(\mathbf{b_i}\), for reward estimation.
Following the linear assumption, the estimated reward is defined as  \(\mathbf{(A_i^{-1}b_i)^\intercal e_t}\) and the confidence interval is defined as \(\alpha\mathbf{\sqrt{e_t^\intercal A_i^{-1} e_t}}\).
The arm that is pulled or the class that is predicted for time \(t\) is the arm that gives the largest sum of the population model probability and the upper confidence bound. 
Providing the probability, \(\mathbf{prob_{t}}\), acts as a method of warm start for the bandits and stabilizes the user experience by starting with more certain predictions.

We refer to a system as any generic system that a user can interact with, as highlighted in Figure \ref{fig:teaser}, where a reward, \(r_{t} \in \{-1, 1\}\), can be inferred from the user interactions: \(r_{t} = 1\) if the previous gesture advances the game, and \(r_{t} = -1\) if the user reports that the gesture was undetected.
The specific system for this paper is the 2-D navigational game described in Section \ref{User Study}. 
While the system could contain the correct labels for each gesture received, the algorithm only needs the sparse reward as a learning signal. 
The system update from the game provides information on either the character advancing in the path or the user feedback, if any, to generate a learning signal.

\subsection{Pre-processing} \label{Pre-processing}
The raw EMG channels were sampled from the sEMG device at 2KHz. 
Pre-processing included bringing the magnitude of the data to a normalized range and removing low-frequency trends with a 40 Hz high-pass filter. 
Acceleration data from an inertial measurement unit on the same sEMG device, sampled at 100 Hz, was also utilized for the population model.
The population model detects the start and end of gestures from EMG and IMU input.
This model has a similar architecture to the gesture model described in \cite{ctrl2024generic}, employing temporal convolution layers and LSTM layers.
We take the latent embeddings before the classification layer as input to our algorithm. 
Thus, the population model acts as a decoder for the EMG signals and is frozen during usage.

\subsection{Post-processing} \label{Post-processing}
We propose a post-processing step in the form of a threshold algorithm to smooth over predictions and provide a better user experience. 
Sliding windows of size \(W\) over features are used to predict a gesture, implemented as buffers in Algorithm \ref{alg:sub}, which lays out this step in more detail.
We determined that a \(W\) equivalent to 40 ms of EMG data is the proper window size, as this is the average expected time to perform a gesture from previous data collection efforts.
Additionally, we pass the latent embedding through the frozen classification layer to obtain each gesture's probability.
The summed probability across all classes is used as a binary hand activity detector.
Alternatively, the binary activity detection can also come from the system or another sensing modality.
When a summed probability exceeds the threshold, \(\tau_{b}\), we aggregate the bandit predictions.
When the average prediction is above threshold, \(\tau_{e}\), then the gesture prediction will be sent to the system.

\begin{figure}[t]

    \vfill
    \begin{subfigure}[b]{0.45\textwidth}
        \centering
        \includegraphics[width=\linewidth]{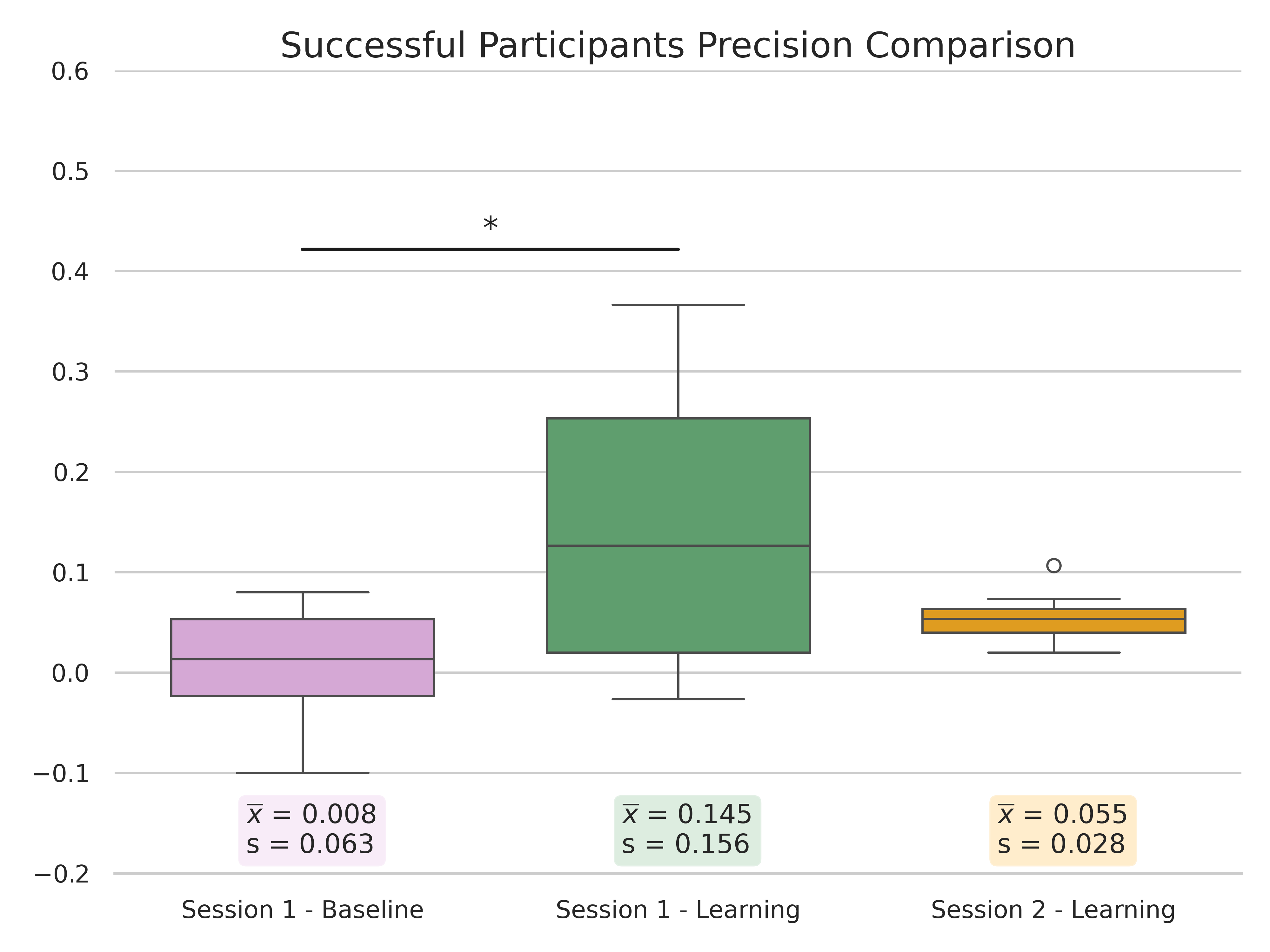}
        \caption{Precision difference in each round for participants able to complete baseline model. There is a significant improvement between the rounds in the first session when learning is enabled.}     
        \captionsetup{justification=centering}
        \label{subfig:prec_non_dnfs}
    \end{subfigure}
    \hfill
    \begin{subfigure}[b]{0.45\textwidth}
        \centering
        \includegraphics[width=\linewidth]{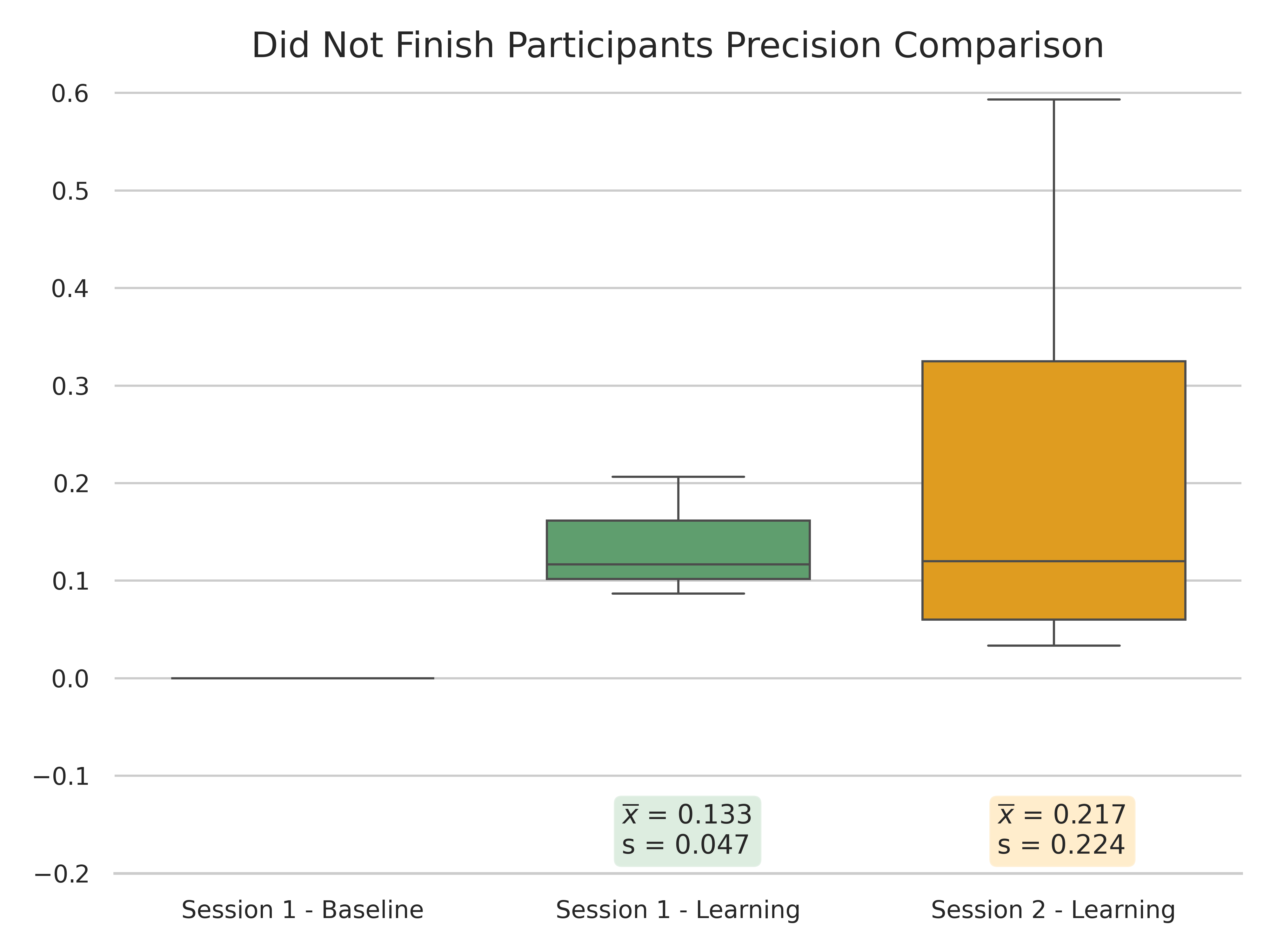}
        \caption{Precision difference in each round for participants unable to complete baseline model. These participants were only able to complete the game with our algorithm enabled.}
        \captionsetup{justification=centering}
        \label{subfig:prec_dnfs}
    \end{subfigure}
    \caption{Comparison of precision difference between rounds, with K = 25.}
\end{figure}

\subsection{User Study} \label{User Study}

We utilized a multi-session experimental design where participants were asked to play a 2-D navigation game utilizing the sEMG-RD device.
This study was approved by the Institutional Review Board (IRB) under Advarra IRB and was conducted in accordance with the principles outlined in the Declaration of Helsinki. 
All participants provided written informed consent prior to participating in the study, and were fully debriefed upon completion of the experiment. 
The game was introduced to each participant by a research assistant, and the participant went through a dry run before the experiment began.
Participants moved the character with thumb swipes, moving their thumb above or on the index finger in the four directions: up, down, left, right. 
At certain points during navigation, participants would perform a specific action rather than a swipe to proceed.
The specific action was either an index pinch or a thumb tap.
An index pinch is defined as tapping the tip of the thumb to the tip of the index finger, and a thumb tap is defined as tapping the thumb on the second joint of the index finger.
Paths consisting of gestures were pseudo-randomly generated with a fixed length at the beginning of each round.
If the participant was successful in advancing the character, this system change is sent to the algorithm as binary inferred feedback or a positive reward to learn from.
Participants were asked to press the spacebar if the character did not respond to their gesture.
This is considered binary user feedback or a negative reward for the algorithm.
The game interface is shown in \ref{fig:teaser}, with a Pac-Man character and grey pellets indicating the path.
At certain grey pellets, an action pop-up indicates to the user to do an action rather than a navigation swipe.

Participants evaluated in this study completed two sessions, with approximately one week between sessions.
During the first session, participants completed the game with the population model for a baseline, then they used our method for the second round.
This is referred to as the Session 1 - Baseline and Session 1 - Learning respectively.
When participants returned for the second session, they continued with their personalized model for one final round of the game, referred to as Session 2 - Learning.

In total, 14 participants (7 male, 7 female, ages 19-58, various occupational backgrounds) were recruited and completed both sessions. 
All participants were right-hand dominant, except for one who was ambidextrous, and performed the gestures with their right hands.
From the exit survey, the participants self-reported their frequency of playing video games on the Likert scale (1-5), with a mean of 3.071 and standard deviation of 0.961.
Similarly for the frequency of using augmented reality or virtual reality devices, the participants reported on the Likert scale (1-5), with a mean of 1.929 and standard deviation of 0.593.
No participants had previous experience with an EMG device.
We also measured participant sentiment with an exit survey, with questions on the Likert scale (1-7).
\begin{figure}[t]

    \centering
    \includegraphics[width=\linewidth]{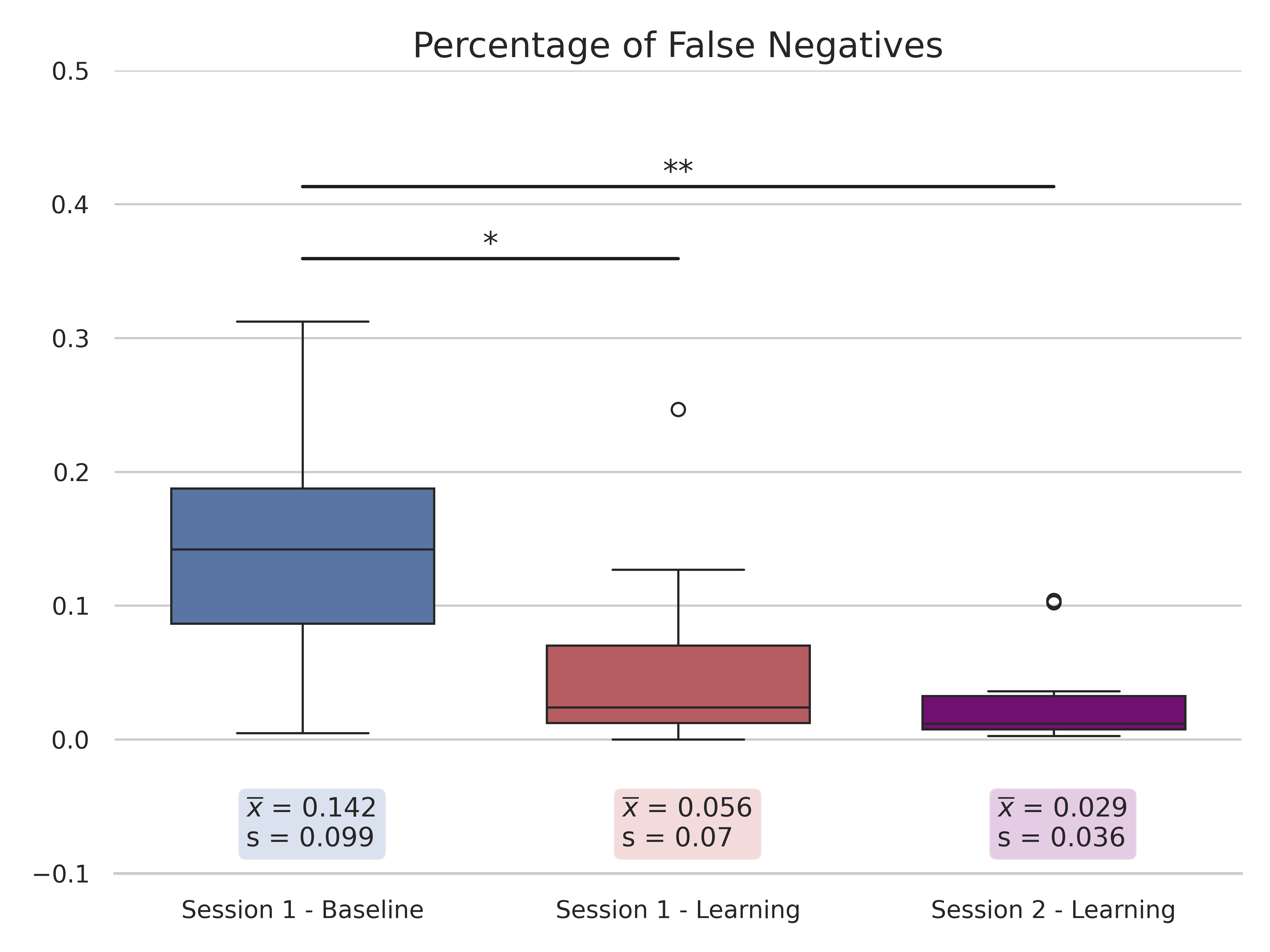}
    \caption{Comparison of percentage of false negatives over all actions between rounds. There is significant decrease with our algorithm enabled.} 
    \captionsetup{justification=centering}
    \label{fig:fn_all}

\end{figure}
\section{Results}

\subsection{Participant Performance}

The average performance across all participants and gestures is illustrated in Figure \ref{fig:average_performance}.
Each data point is calculated on the aggregate of five trials to ensure a balance of gestures.
The first curve from the baseline round in the first session has an initial uptick of improvement.
This is most likely due to an initial human learning on the part of the user rather than the model, because the model was static.
When utilizing the proposed bandit model, the latter two rounds show the longitudinal positive benefit of personalization.

To measure the improvement through a round, we look at the performance of each participant's first and last \(K=25\) attempts of each gesture, coming approximately from the first and last 10-12 trials of a round.
We look at the difference in precision of the last \(K\) and first \(K\) of each gesture as a metric to determine the amount of benefit our method provides.
There were a subsert of participants who were unable to complete the baseline model, which we refer to as the Did Not Finish participants.  
These participants were only able to complete the game during the next round (Session 1 - Learning) with the bandits algorithm.

Following Figure \ref{subfig:prec_non_dnfs}, participants successful in completing the baseline round achieved an average improvement of 0.008 between the start and end of the baseline. 
However, they saw an average increase of 0.145 in the next round (Session 1 - Learning) with the proposed algorithm.
Using a paired sample T-Test, we found that there is a significant improvement from the baseline round, Session 1 - Baseline and Session 1 - Learning, \(T(df=6), p=0.0288<0.05\)
This indicates that there is a clear performance benefit when using the bandits algorithm over the baseline.
While the Did Not Finish participants were unable to complete the baseline for a similar analysis, they were only able to complete the game with our method and ultimately, saw an average improvement of 0.133 in Session 1 - Learning, as shown in Figure \ref{subfig:prec_dnfs}.
The Did Not Finish participants continued to achieve a larger increase of precision in the second session (Session 2 - Learning) for an improvement of 0.217 on average.

We were able to approximate the false negatives from the spacebar presses initiated by the participant.
We measure the percentage of false negatives to total actions performed in each round to look at the impact of the proposed model on false negatives.
Figure \ref{fig:fn_all} shows the results.
Utilizing a paired sample T-test, we found that there is a significant decrease in the percentage from Session 1 - Baseline to Session 1 - Learning, \(T(df=13), p=0.0233<0.05\).
Similarly, between Session 1 - Baseline and Session 2 - Learning, we found that there is a significant decrease, \(T(df=13), p=0.002<0.01\).
These findings indicate that the bandits algorithm significantly reduced the percentage of false negatives as the algorithm personalized to participants.
\subsection{Participant Sentiment}
\begin{figure}[t]
    \vfill
    \begin{subfigure}[b]{0.45\textwidth}
        \centering
        \includegraphics[width=\linewidth]{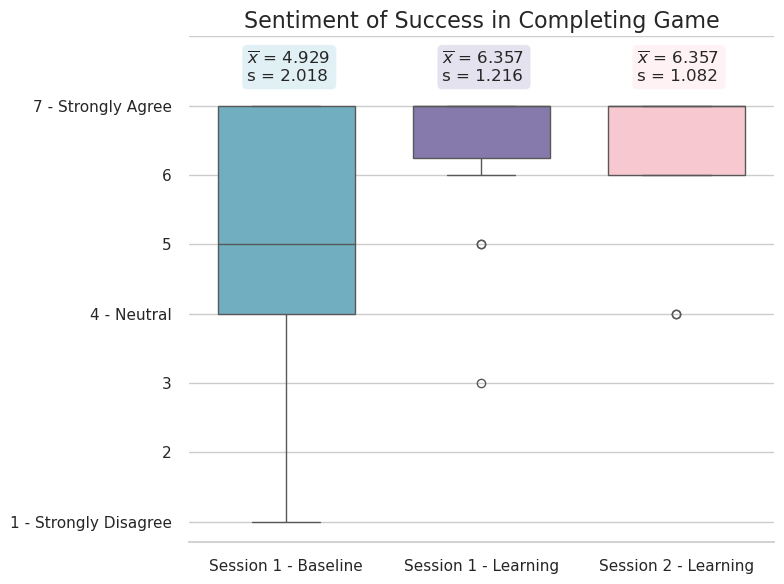}
        \caption{Self reported feelings of success in finishing the game after each round. With our algorithm, participants reported feeling more successful.}      
        \captionsetup{justification=centering} 
        \label{subfig:success}
    \end{subfigure}
    \hfill
    \begin{subfigure}[b]{0.45\textwidth}
        \centering
        \includegraphics[width=\linewidth]{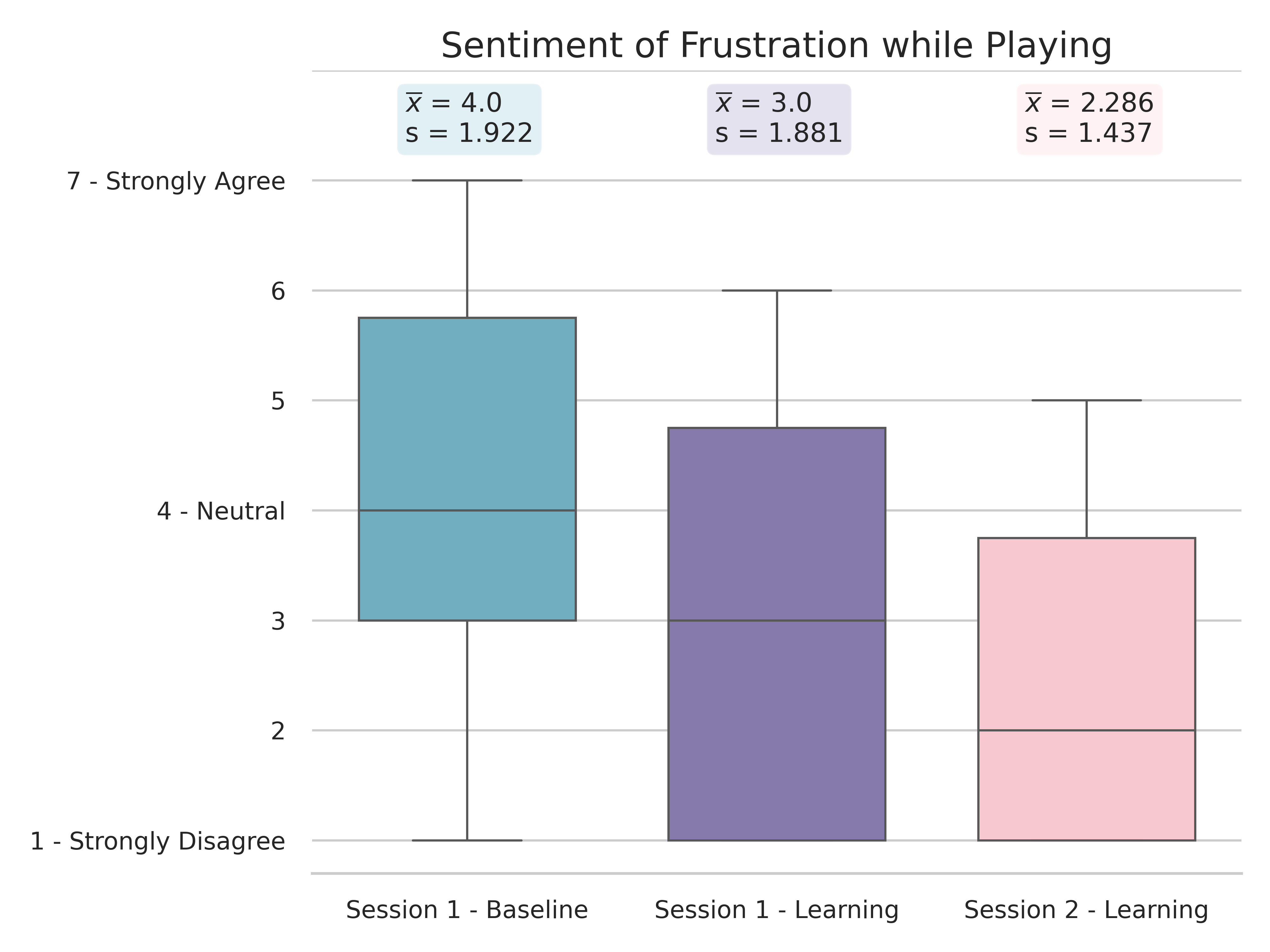}
        \caption{Self reported feelings of frustration while playing the game after each round. With our algorithm, participants reported feeling less frustrated.}       
        \captionsetup{justification=centering}
        \label{subfig:frustration}
    \end{subfigure}
    \caption{Comparison of user sentiment between rounds.}
    \label{fig:sentiment}
    
\end{figure}

To gain a deeper understanding of the user experience with an online personalization algorithm, we measured participant sentiment after each round.
In the exit surveys, participants expressed their sentiments regarding success in completing the game and frustration while playing the game after each round.
Participants showed a clear decrease in frustration as the users progressed to the later rounds, as illustrated in Figure \ref{subfig:frustration}.
Similarly, participants reported feeling that they were more successful in completing the games in the rounds using the bandits algorithm than the baseline as shown in Figure \ref{subfig:success}.
By the end of engaging with the learning algorithm, the majority of participants agreed that they were successful in completing the game.
Together, the participant sentiments show that compared to the initial responses in Session 1 - Baseline, the user experience increased with the learning algorithm and continued to increase with more time.

\section{Conclusion}

In this work, we explore the application of reinforcement learning for online longitudinal personalization of a gesture recognition algorithm.
Applying this method with a sEMG-RD device allowed us to conduct a user study, demonstrating not only the improvements in precision by using the bandit algorithm but also enabling a subset of users to successfully finish the game.
We show that the algorithm can leverage implicit signals from a system and explicit signals from the user to personalize a hand gesture recognition model, without incurring additional false negatives or deteriorating user experience.

In future work, researchers may explore in more detail the algorithm's linear assumption and the derivation of rewards.
There are limitations in the linear assumption between the reward estimate and the features.
\cite{zhou2019neuralucb} proposes a non-linear approximation of the reward and estimates the uncertainty from a neural network.
This formulation, while providing a more expressive reward estimation, requires more data initially to learn meaningful embeddings in the neural network.
Future efforts can look at combining learned pre-trained representations such as the population embeddings in this work and more complex upper confidence bound formulations.
Additionally, the system used in this study provides easily derived positive reward signals from the game progression.
However, the negative rewards from the undetected gestures or false negatives rely solely user feedback.
Understanding the balance of negative and positive rewards in future work would be an interesting direction to inform designs of new reward systems.
Alternate systems can derive reward signals from other sensing modalities that can confirm a successful prediction as a positive reward or from a well-designed user interface with limited state transitions such that any erroneous transition can be used as a negative reward.
The topic of developing future systems capable of providing reward signals is left to future work.

\section{Acknowledgements}
We thank Casey Brown and Alexander Bakogeorge for user study support, and Zhenxing Han for software support, and Emily Mugler, Frances Lau, Hrvoje Benko for input and discussion. 

\bibliographystyle{plainnat}
\bibliography{base}

\begin{thebibliography}{15}
\providecommand{\natexlab}[1]{#1}
\providecommand{\url}[1]{\texttt{#1}}
\expandafter\ifx\csname urlstyle\endcsname\relax
  \providecommand{\doi}[1]{doi: #1}\else
  \providecommand{\doi}{doi: \begingroup \urlstyle{rm}\Url}\fi

\bibitem[Akbari et~al.(2021)Akbari, Martinez, and Jafari]{akbari2021meta}
Ali Akbari, Jonathan Martinez, and Roozbeh Jafari.
\newblock A meta-learning approach for fast personalization of modality
  translation models in wearable physiological sensing.
\newblock \emph{IEEE journal of biomedical and health informatics}, 26\penalty0
  (4):\penalty0 1516--1527, 2021.

\bibitem[Berman et~al.(2024)Berman, Lee, Yin, and Huang]{berman2024efficient}
Joseph Berman, I-Chieh Lee, Jie Yin, and He~Huang.
\newblock An efficient framework for personalizing emg-driven musculoskeletal
  models based on reinforcement learning.
\newblock \emph{IEEE Transactions on Neural Systems and Rehabilitation
  Engineering}, 2024.

\bibitem[Chen et~al.(2023)Chen, Wang, Quan, Peng, Lin, Srivastava, Matusik, and
  Stankovic]{chen2023robust}
Wenqiang Chen, Ziqi Wang, Pengrui Quan, Zhencan Peng, Shupei Lin, Mani
  Srivastava, Wojciech Matusik, and John Stankovic.
\newblock Robust finger interactions with cots smartwatches via unsupervised
  siamese adaptation.
\newblock In \emph{Proceedings of the 36th Annual ACM Symposium on User
  Interface Software and Technology}, pages 1--14, 2023.

\bibitem[Den~Hengst et~al.(2020)Den~Hengst, Grua, el~Hassouni, and
  Hoogendoorn]{den2020reinforcement}
Floris Den~Hengst, Eoin~Martino Grua, Ali el~Hassouni, and Mark Hoogendoorn.
\newblock Reinforcement learning for personalization: A systematic literature
  review.
\newblock \emph{Data Science}, 3\penalty0 (2):\penalty0 107--147, 2020.

\bibitem[Despradel et~al.(2024)Despradel, Murphy, Borda, Verma, Yadav,
  Shanahan, Marshall, Formento, Bracklein, Ye, et~al.]{despradel2024enabling}
Dailyn Despradel, Max Murphy, Luigi Borda, Nikhil Verma, Prakarsh Yadav, Jenn
  Shanahan, Najja Marshall, Emanuele Formento, Mario Bracklein, Jun Ye, et~al.
\newblock Enabling advanced interactions through closed-loop control of motor
  unit activity after tetraplegia.
\newblock In \emph{Adjunct Proceedings of the 37th Annual ACM Symposium on User
  Interface Software and Technology}, pages 1--3, 2024.

\bibitem[Joshi et~al.(2017)Joshi, Ghosh, Betke, Sclaroff, and
  Pfister]{joshi2017personalizing}
Ajjen Joshi, Soumya Ghosh, Margrit Betke, Stan Sclaroff, and Hanspeter Pfister.
\newblock Personalizing gesture recognition using hierarchical bayesian neural
  networks.
\newblock In \emph{Proceedings of the IEEE Conference on Computer Vision and
  Pattern Recognition}, pages 6513--6522, 2017.

\bibitem[labs~at Reality~Labs et~al.(2024)labs~at Reality~Labs, Sussillo,
  Kaifosh, and Reardon]{ctrl2024generic}
Ctrl labs~at Reality~Labs, David Sussillo, Patrick Kaifosh, and Thomas Reardon.
\newblock A generic noninvasive neuromotor interface for human-computer
  interaction.
\newblock \emph{bioRxiv}, pages 2024--02, 2024.

\bibitem[Li et~al.(2010)Li, Chu, Langford, and Schapire]{Li_2010}
Lihong Li, Wei Chu, John Langford, and Robert~E. Schapire.
\newblock A contextual-bandit approach to personalized news article
  recommendation.
\newblock In \emph{Proceedings of the 19th international conference on World
  wide web}, WWW ’10, page 661–670. ACM, April 2010.
\newblock \doi{10.1145/1772690.1772758}.
\newblock URL \url{http://dx.doi.org/10.1145/1772690.1772758}.

\bibitem[Lingler et~al.(2024)Lingler, Talypova, Jokinen, Oulasvirta, and
  Wintersberger]{10.1145/3613904.3642063}
Alexander Lingler, Dinara Talypova, Jussi P.~P. Jokinen, Antti Oulasvirta, and
  Philipp Wintersberger.
\newblock Supporting task switching with reinforcement learning.
\newblock In \emph{Proceedings of the 2024 CHI Conference on Human Factors in
  Computing Systems}, CHI '24, New York, NY, USA, 2024. Association for
  Computing Machinery.
\newblock ISBN 9798400703300.
\newblock \doi{10.1145/3613904.3642063}.
\newblock URL \url{https://doi.org/10.1145/3613904.3642063}.

\bibitem[Lu et~al.(2024)Lu, Chen, Hsu, Deshpande, Wang, and
  MacIntyre]{10.1145/3613905.3651059}
Feiyu Lu, Mengyu Chen, Hsiang Hsu, Pranav Deshpande, Cheng~Yao Wang, and Blair
  MacIntyre.
\newblock Adaptive 3d ui placement in mixed reality using deep reinforcement
  learning.
\newblock CHI EA '24, New York, NY, USA, 2024. Association for Computing
  Machinery.
\newblock ISBN 9798400703317.
\newblock \doi{10.1145/3613905.3651059}.
\newblock URL \url{https://doi.org/10.1145/3613905.3651059}.

\bibitem[Madduri et~al.(2024)Madduri, Yamagami, Li, Burckhardt, Burden, and
  Orsborn]{madduri2024predicting}
Maneeshika~M Madduri, Momona Yamagami, Si~Jia Li, Sasha Burckhardt, Samuel~A
  Burden, and Amy~L Orsborn.
\newblock Predicting and shaping human-machine interactions in closed-loop,
  co-adaptive neural interfaces.
\newblock \emph{bioRxiv}, pages 2024--05, 2024.

\bibitem[Rahimian et~al.(2021)Rahimian, Zabihi, Asif, Atashzar, and
  Mohammadi]{rahimian2021few}
Elahe Rahimian, Soheil Zabihi, Amir Asif, S~Farokh Atashzar, and Arash
  Mohammadi.
\newblock Few-shot learning for decoding surface electromyography for hand
  gesture recognition.
\newblock In \emph{ICASSP 2021-2021 IEEE International Conference on Acoustics,
  Speech and Signal Processing (ICASSP)}, pages 1300--1304. IEEE, 2021.

\bibitem[Shen et~al.(2024)Shen, Xu, Tan, Karlson, and
  Strasnick]{shen2024boosting}
Junxiao Shen, Xuhai Xu, Ran Tan, Amy Karlson, and Evan Strasnick.
\newblock Boosting gesture recognition with an automatic gesture annotation
  framework.
\newblock In \emph{2024 IEEE 18th International Conference on Automatic Face
  and Gesture Recognition (FG)}, pages 1--10. IEEE, 2024.

\bibitem[Xu et~al.(2022)Xu, Gong, Brum, Liang, Suh, Gupta, Agarwal, Lindsey,
  Kang, Shahsavari, et~al.]{xu2022enabling}
Xuhai Xu, Jun Gong, Carolina Brum, Lilian Liang, Bongsoo Suh, Shivam~Kumar
  Gupta, Yash Agarwal, Laurence Lindsey, Runchang Kang, Behrooz Shahsavari,
  et~al.
\newblock Enabling hand gesture customization on wrist-worn devices.
\newblock In \emph{Proceedings of the 2022 CHI Conference on Human Factors in
  Computing Systems}, pages 1--19, 2022.

\bibitem[Zhou et~al.(2019)Zhou, Li, and Gu]{zhou2019neuralucb}
Dongruo Zhou, Lihong Li, and Quanquan Gu.
\newblock Neural contextual bandits with upper confidence bound-based
  exploration.
\newblock \emph{CoRR}, abs/1911.04462, 2019.
\newblock URL \url{http://arxiv.org/abs/1911.04462}.

\end{thebibliography}

\end{document}